\documentstyle[prl,twocolumn,aps,amsfonts]{revtex}
\def\DSAS{\Delta_{\text{SAS}}}
\def\HilbLLL{\Bbb{H}_{\text{LLL}}}
\def\spin{\text{spin}}

\def\sky{\text{sky}}
\def\eff{\text{eff}}
\def\dx{d^{2}x}
\def\dy{d^{2}y}
\def\B{{\cal B}}
\def\upA{\uparrow}
\def\dnA{\downarrow}
\def\a{{\cal A}}

\def\P{{\cal P}}

\def\bpmatrix{\left(\begin{array}{c}}
\def\epmatrix{\end{array}\right)}
\def\F{{\frak S}}
\def\Laugh{\F_{\text{LN}}[\bbox{x}]}
\def\txg{\text{g}}
\input epsf
\begin{document}
\title{Spin-Pseudospin Coherence and CP$^{3}$ Skyrmions \\
in Bilayer Quantum Hall Ferromagnets
}
\author{Z. F. Ezawa}
\address{
Department of Physics, Tohoku University, Sendai 980, Japan
}
\maketitle\begin{abstract}
We analyze bilayer quantum Hall ferromagnets, whose underlying symmetry group 
is SU(4).  Spin-pseudospin coherence develops spontaneously when the total 
electron density is low enough.  Quasiparticles are CP$^{3}$ skyrmions.  One 
skyrmion induces charge modulations on both of the two layers.  At the filling 
factor$\nu =2/m$ one elementary excitation consists of a pair of skyrmions and 
its charge is $2e/m$.  Recent experimental data due to Sawada et al. 
[Phys. Rev. Lett. {\bf 80}, 4534 (1998)] support this conclusion.
\newline
\end{abstract}

The quantum Hall (QH) effect is a remarkable macroscopic quantum 
phenomenon in the two-dimensional electron system \cite{FQHEbook}.  Attention has 
recently been paid to quantum coherence in QH systems.  The kinetic and 
Coulomb Hamiltonians have the spin SU(2) symmetry.  Its spontaneous breakdown 
leads to a spin coherence and turns the system into a QH ferromagnet.  The 
effective Hamiltonian is the SU(2) nonlinear sigma (NL$\sigma $) model \cite{Dadda}.  
Quasiparticles are CP$^{1}$ skyrmions \cite{SkyrmQH,Barrett,Schmeller}.  

In this paper we analyze skyrmion excitations in bilayer QH (BLQH) 
ferromagnets.  The lowest Landau level (LLL) contains four energy levels 
corresponding to the spin and layer (pseudospin) degrees of freedom.  The 
SU(4) symmetry underlies the BLQH system.  Its spontaneous breakdown leads to 
a spin-pseudospin coherence.  The effective Hamiltonian is the SU(4) NL$\sigma $ 
model \cite{Dadda}.  Quasiparticles are CP$^{3}$ skyrmions \cite{Dadda}.  They have two 
characteristic features: (A) One skyrmion induces charge modulations on both 
of the two layers.  The main part of the activation energy is the capacitive 
charging energy.  (B) One elementary excitation consists of a pair of 
skyrmions at $\nu =2/m$ with $m$ odd.  It carries the charge $2e/m$.  Our 
theoretical analysis accounts for recent experimental data due to Sawada et 
al.\cite{SawaPRLa,SawaPRLb}.  Throughout the paper we use the natural units 
$\hbar =c=1$.

\textit{QH ferromagnets}: 
We analyze skyrmion excitations at the filling factor $\nu \equiv 2\pi \rho _{0}/eB_{\perp }=1/m$ with 
$m$ odd.  We use an improved composite-boson (CB) theory \cite{EzaICBa}, which is 
proposed based on a suggestion due to Girvin et al. \cite{GirvinAAA}. The 
advantage of the scheme is a direct connection between the semiclassical 
property of an excitation and its microscopic wave function.  We start with a 
review of monolayer QH ferromagnets \cite{EzaICBa}.  The analysis of BLQH 
ferromagnets is its straightforward generalization with a replacement of SU(2) 
by SU(4).  

The kinetic Hamiltonian for planar electrons with mass $M$ in the 
perpendicular magnetic field $B_{\perp }$ is
\begin{equation}
H_{K}= {1\over 2M}\int \dx \Psi ^{\dagger }(\bbox{x})(P_{x}-iP_{y})(P_{x}+iP_{y})\Psi (\bbox{x}) ,
\label{HamilKinet}
\end{equation}
where $P_{j}=-i\partial _{j}+eA_{j}$ is the covariant momentum with $A_{j}={1\over 2}\varepsilon _{jk}x_{k}B_{\perp }$; $\varepsilon _{12}=-
\varepsilon _{21}=1$ and $\varepsilon _{11}=\varepsilon _{22}=0$.  The electron field $\Psi (\bbox{x})$ is a two-component spinor 
made of the spin-up ($\psi ^{\upA }$) and spin-down ($\psi ^{\dnA }$) field.  The symmetry group of 
this Hamiltonian is U(2)=U(1)$\otimes $SU(2).

When the Zeeman energy is small, a spin coherence develops 
spontaneously.  This is described by introducing the two-component CB field by 
the formula \cite{EzaICBa,GirvinAAA},
\begin{equation}
\Phi (\bbox{x})=e^{-\a(\bbox{x})}e^{-ie\Theta (\bbox{x})}\Psi (\bbox{x}) , 
\label{DressField}
\end{equation}
where $\a(\bbox{x})$ is the auxiliary field obeying $\bbox{\nabla }^{2}\a(\bbox{x})=2\pi m[\rho (\bbox{x})-\rho _{0}]$; 
$\rho (\bbox{x})\equiv \Psi ^{\dagger }(\bbox{x})\Psi (\bbox{x})$ is the electron density.  The phase field $\Theta (\bbox{x})$ attaches $m$ 
units of flux quantum $2\pi /e$ to each electron via the relation, 
$\varepsilon _{ij}\partial _{i}\partial _{j}\Theta (\bbox{x})=(2\pi /e)m\rho (\bbox{x})$.  The effective magnetic field is
\begin{equation}
\B_{\eff}(\bbox{x})=B_{\perp } - \varepsilon _{ij}\partial _{i}\partial _{j}\Theta (\bbox{x}) = B_{\perp }-(2\pi /e)m\rho (\bbox{x}) .
\label{EffecMagne}
\end{equation}
It vanishes, $\langle \B_{\eff}\rangle _{\txg}=0$, on the ground state at $\nu =1/m$.  Substituting 
(\ref{DressField}) into (\ref{HamilKinet}), the kinetic Hamiltonian reads
\begin{equation}
H_{K} = {1\over 2M}\int d^{2}x \Phi ^{\ddag }(\bbox{x})(\P_{x}-i\P_{y})(\P_{x}+i\P_{y})\Phi (\bbox{x}) ,
\label{HamilCB}
\end{equation}
where $\P_{j} =  -i\partial _{j} - (\varepsilon _{jk} + i\delta _{jk})\partial _{k}\a(\bbox{x})$ is the covariant momentum.  We have 
defined $\Phi ^{\ddag }(\bbox{x})\equiv \Phi ^{\dagger }(\bbox{x})e^{2\a(\bbox{x})}$, with which $\rho (\bbox{x})=\Psi ^{\dagger }(\bbox{x})\Psi (\bbox{x})=\Phi ^{\ddag }(\bbox{x})\Phi (\bbox{x})$.  An 
analysis of the Lagrangian shows that the canonical conjugate of $\varphi ^{\alpha }(\bbox{x})$ is 
not $i\varphi ^{\alpha \dagger }(\bbox{x})$ but $i\varphi ^{\alpha \ddag }(\bbox{x})$.  

We decompose the CB field as
\begin{equation}
\Phi (\bbox{x}) = e^{-\a(\bbox{x})}\phi (\bbox{x})\bbox{n}(\bbox{x}) ,
\label{FieldCP}
\end{equation}
with the U(1) component $\phi (\bbox{x})=e^{i\chi (\bbox{x})}\sqrt {\rho (\bbox{x})}$ and the SU(2) component $\bbox{n}(\bbox{x})$: 
It is the CP$^{1}$ field \cite{Dadda} subject to the constraint, $\bbox{n}^{\dagger }(\bbox{x})\bbox{n}(\bbox{x})=1$.  The 
spin density is 
\begin{equation}
S^{a}(\bbox{x})={1\over 2}\rho (\bbox{x})s^{a}(\bbox{x}),  \quad  s^{a}(\bbox{x})=\bbox{n}^{\dagger }(\bbox{x})\lambda ^{a}\bbox{n}(\bbox{x}) ,
\label{SpinDensi}
\end{equation}
where $\lambda ^{a}$ are the Pauli matrices.

At sufficiently low temperature it is reasonable to focus our attention 
to physics taking place within the LLL.  The Hilbert space $\HilbLLL$ is 
constructed by imposing the LLL condition so that the kinetic term (\ref{HamilCB}) 
is quenched.  It has a simple expression in terms of the CB field,
\begin{equation}
(\P_{x}+i\P_{y})\Phi (\bbox{x})|\F\rangle =-{i\over \ell _{B}}{\partial \over \partial z^{*}}\Phi (\bbox{x})|\F\rangle  = 0 .
\label{LLLcondiDress}
\end{equation}
The complex number is $z=(x+iy)/2\ell _{B}$ with $\ell _{B}$ the magnetic length.  Hence, 
the wave function for composite bosons is analytic and symmetric in all $N$ 
coordinates,
\begin{equation}
\Omega [z] = \langle 0|\Phi (\bbox{x}_{1})\cdots \Phi (\bbox{x}_{N})|\F\rangle  .
\label{WaveFunctDress}
\end{equation}
The wave function for electrons is $\F[\bbox{x}]=\Omega [z]\Laugh$, where $\Laugh$ is the 
familiar Laughlin wave function.  Here, $[\bbox{x}]=(\bbox{x}_{1},\bbox{x}_{2},\cdots ,\bbox{x}_{N})$ and 
$[z]=(z_{1},z_{2},\cdots ,z_{N})$.  Any excitation confined to the LLL is described by a 
choice of the analytic spinor factor $\Omega [z]$.

We analyze the CB theory semiclassically, where the bosonic field 
operator is approximated by a c-number field.  It follows from 
(\ref{WaveFunctDress}) that the wave function is
\begin{equation}
\F[\bbox{x}] = \prod _{r}\langle \Phi (\bbox{x}_{r})\rangle \Laugh, 
\label{FactoWave}
\end{equation}
where $\langle \Phi (\bbox{x})\rangle $ is analytic.  If the Zeeman energy is neglected, the ground 
state is determined by minimizing the Coulomb energy.  The one-point function 
$\langle \Phi (\bbox{x})\rangle $ is a constant vector pointing to an arbitrary direction in the SU(2) 
space, as implies a spontaneous breakdown of the SU(2) symmetry.  Actually, a 
small Zeeman interaction fixes this direction so that $\langle s^{a}(\bbox{x})\rangle =\delta ^{az}$, or
\begin{equation}
\langle \varphi ^{\upA }(\bbox{x})\rangle _{\txg}= \sqrt {\rho _{0}}, \quad  \langle \varphi ^{\dnA }(\bbox{x})\rangle _{\txg}= 0.
\label{GrounSpin}
\end{equation}
The ground-state wave function is given by (\ref{FactoWave}) with (\ref{GrounSpin}).

The semiclassical LLL condition follows from (\ref{FieldCP}),
\begin{equation}
\langle \varphi ^{\alpha }(\bbox{x})\rangle = e^{-\a(\bbox{x})}e^{i\chi (\bbox{x})} \sqrt {\rho (\bbox{x})}n^{\alpha }(\bbox{x}) \equiv  \omega ^{\alpha }(z),
\label{PreSolitEq}
\end{equation}
where various fields are classical ones.  This is solved by $\chi (\bbox{x})=0$ and 
$n^{\alpha }(\bbox{x})=\omega ^{\alpha }(z)/\sqrt {\sum _{\alpha }|\omega ^{\alpha }|^{2}}$.  The soliton equation \cite{EzaIQC} follows trivially 
from (\ref{PreSolitEq}),
\begin{equation}
{\nu \over 4\pi }\bbox{\nabla }^{2}\ln \rho (\bbox{x}) - \rho (\bbox{x}) + \rho _{0}= \nu Q(\bbox{x}) ,
\label{SolitEq}
\end{equation}
where $Q(\bbox{x})$ is the topological charge density,
\begin{equation}
Q(\bbox{x}) = {1\over 4\pi }\bbox{\nabla }^{2}\ln\bigl(\sum _{\alpha }|\omega ^{\alpha }(z)|^{2}\bigr).
\label{SkyrmCharg}
\end{equation}
The lightest skyrmion on the ground state (\ref{GrounSpin}) is
\begin{equation}
\langle \varphi ^{\upA }(\bbox{x})\rangle _{\sky}= z\sqrt {\rho _{0}}, \quad  \langle \varphi ^{\dnA }(\bbox{x})\rangle _{\sky}= \kappa \sqrt {\rho _{0}} ,
\label{SkyrmSpin}
\end{equation}
with which the wave function is given by (\ref{FactoWave}).  For a large skyrmion 
($\kappa \gg 1$), the soliton equation (\ref{SolitEq}) is solved iteratively and agrees with 
the familiar expression \cite{SkyrmQH},
\begin{equation}
\varrho  (\bbox{x}) \equiv  \rho (\bbox{x}) - \rho _{0} \simeq  \nu Q(\bbox{x}) = -{\nu \over \pi } {4(\kappa \ell _{B})^{2}\over [r^{2}+4(\kappa \ell _{B})^{2}]^{2}} .
\label{SkyrmDensi}
\end{equation}
The skyrmion is reduced to the vortex in the limit $\kappa \rightarrow 0$.

The excitation energy of one skyrmion consists of the exchange energy 
$E_{\text{ex}}$, the electrostatic energy $E_{C}$ and the Zeeman energy $E_{Z}$.  
Minimizing their sum, we determine the skyrmion size $\kappa $, the skyrmion energy 
$E_{\sky}$ and the flipped-spin number $N_{\spin}$ as follows \cite{EzaICBa}:
\begin{eqnarray}
\kappa &&\simeq  {1\over 2}(\beta \nu )^{1/3}\biggl\{\widetilde{g}\ln\biggl({\sqrt {2\pi }\over 32\widetilde{g}}+1\biggr)\biggr\}^{-1/3} , \label{OptimScale}\\
E_{\sky}&&\simeq  \biggl(\nu \sqrt {{\pi \over 32}}+{3\beta \nu ^{2}\over 4\kappa }\biggr)E_{C}^{0} , \label{SkyrmTotalEnerg}\\ 
N_{\spin} &&\simeq  2\nu \kappa ^{2}\ln\biggl({\sqrt {2\pi }\over 32\widetilde{g}}+1\biggr) . \label{FlippSpin}
\end{eqnarray}
Here, $\rho _{s}=\nu e^{2}/(16\sqrt {2\pi }\varepsilon \ell _{B})$, $E_{C}^{0}=e^{2}/\varepsilon \ell _{B}$ and $\widetilde{g}=g^{*}\mu _{B}B/E_{C}^{0}$.  The parameter 
$\beta $ represents the strength of the Coulomb energy;  it is calculated as 
$\beta =3\pi ^{2}/64$ for a sufficiently large skyrmion in an ideal planar system.  
However, an actual skyrmion is not large and there will also be a correction 
from a finiteness of the layer width.  We use $\beta $ as a phenomenological 
parameter.  We fix it as $\beta =0.24$ by requiring the skyrmion spin 
$N_{\text{spin}}$ to agree with the experimental data due to Barrett et al. 
\cite{Barrett} at $\nu =1$: $N_{\spin}\simeq 3.7$ at $B=7.05$ Tesla, where $\kappa \simeq 1.0$ and $\widetilde{g}\simeq 
0.015$.  Note that $N_{\spin}\simeq 5.3$ ($\kappa \simeq 1.10$) at $B=3$ Tesla and $N_{\spin}\simeq 2.7$  
($\kappa \simeq 0.96$) at $B=15$ Tesla.  Our results are consistent with the previous ones 
\cite{SkyrmQH}.

\begin{figure}[thb]
\epsfxsize=70mm
\epsfbox{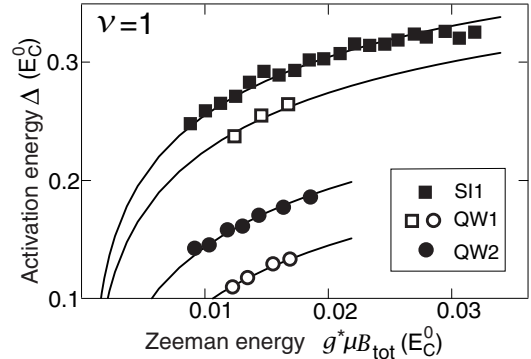}
\caption{
Theoretical curves versus experimental data due to Schmeller et al.[5] for the 
activation energy in three samples SI1, QW1 and QW2.  There are two curves for 
one sample (QW1) but with different mobilities.  The offset $\Gamma _{\text{offset}}$ 
increases as the mobility decreases.  
}
\label{SkyrEneZPS}
\end{figure}

The excitation energy of a skyrmion-antiskyrmion pair will be given by 
$\Delta =2E_{\sky}-\Gamma _{\text{offset}}$ with a sample dependent offset $\Gamma _{\text{offset}}$.  In 
Fig.\ref{SkyrEneZPS} we have fitted the experimental data due to Schmeller et 
al. \cite{Schmeller} based on formula (\ref{SkyrmTotalEnerg}) with $\beta =0.24$, where an 
appropriate offset $\Gamma _{\text{offset}}$ is used for each curve.  The theoretical 
curves reproduce all data remarkably well.

\textit{BLQH systems}:  We proceed to analyze BLQH states at $\nu =1/m$ 
and $2/m$ with $m$ odd.  There are three experimental techniques to elucidate 
various states.  The first one is to change the total electron density $\rho _{0}$.  
By increasing $\rho _{0}$ the interlayer separation $d$ is effectively increased 
compared with the magnetic length, $d/\ell _{B}\propto \sqrt {\rho _{0}}$.  Hence, as $\rho _{0}\rightarrow \infty $ the BLQH 
system at $\nu =2/m$ will be decomposed into two independent $\nu =1/m$ monolayer 
systems;  as $\rho _{0}\rightarrow 0$ an interlayer coherence will develop spontaneously, which 
has been argued \cite{EIcoher,EzaIQC,Moon} at $\nu =1/m$ and will be argued at 
$\nu =2/m$ in this paper.  Consequently, we expect a phase transition at $\nu =2/m$ 
between these two phases but not at $\nu =1/m$.  These two phases are clearly 
distinguishable by using the second technique, i.e. by applying gate bias 
voltage.  We can control the density difference $\sigma _{0}$ between the two quantum 
wells, where $\sigma _{0}=(\rho _{0}^{1}-\rho _{0}^{2})/\rho _{0}$ with $\rho _{0}^{\alpha }$ the density in the layer $\alpha $.  Only 
the coherent state is stable \cite{EIcoher} against an arbitrary change of $\sigma _{0}$.   
All these features have been experimentally confirmed in recent works due to 
Sawada et al. \cite{SawaPRLa} at $\nu =1$ and $2$.  The coherent state at $\nu =2/3$ has 
not been observed by them presumably because of a poor sample quality.  The 
third technique is to tilt the sample with the perpendicular magnetic field 
$B_{\perp }$ fixed.  In the high-density data \cite{SawaPRLb} the activation energy is 
found to increase at $\nu =2$ and $2/3$ as normally as in the monolayer system: 
Indeed, we can fit the data by the monolayer skyrmion formula 
(\ref{SkyrmTotalEnerg}).  We conclude that elementary excitations are monolayer 
skyrmions.  In the low-density data \cite{SawaPRLb} it is found to decrease 
anomalously at $\nu =1$ and $\nu =2$, as is the phenomenon discovered by Murphy et 
al. \cite{Sheena} at $\nu =1$:  It is a behavior intrinsic to the interlayer coherent 
state in the BLQH system.

\textit{Spin-pseudospin coherence}: 
We analyze elementary excitations in the coherent state of the BLQH system.  
The electron field $\Psi (\bbox{x})$ has four components $\psi ^{1\upA }$, $\psi ^{1\dnA }$, $\psi ^{2\upA }$ and $\psi ^{2\dnA }$, 
where the layer is indexed by 1 and 2.  The kinetic Hamiltonian is given by 
(\ref{HamilKinet}), whose symmetry group is U(4)=U(1)$\otimes $SU(4).  When the interlayer 
and intralayer Coulomb energies are nearly equal and dominate the system, we 
expect a spin-pseudospin coherence to emerge.  

Such a new phase is described in terms of the CB field defined by 
(\ref{DressField}).  The CB field is decomposed into the U(1) and SU(4) components 
by (\ref{FieldCP}).  Here, $\bbox{n}(\bbox{x})$ is the CP$^{3}$ field.  The group SU(4) is generated 
by the hermitian, traceless, $4\times 4$ matrices $\lambda ^{a}$, $a=1,2,\cdots ,15$, normalized as 
$\text{Tr}(\lambda ^{a}\lambda ^{b})=2\delta ^{ab}$.  They are the generalization of the Pauli matrices in 
case of SU(2).  The SU(4) spin density is given by (\ref{SpinDensi}) with such 
$\lambda ^{a}$.

The Coulomb energy is decomposed into two terms,
\begin{equation}
E_{C}^{\pm } = {1\over 2}\int \dx\dy V^{+}(\bbox{x}-\bbox{y})\varrho  _{\pm }(\bbox{x})\varrho  _{\pm }(\bbox{y}), \label{HamilCouloBL}
\end{equation}
where $V^{\pm }(\bbox{x})=(e^{2}/2\varepsilon )\bigl(|\bbox{x}|^{-1} \pm  (|\bbox{x}|^{2}+d^{2})^{-1/2}\bigr)$ with the interlayer separation 
$d$; $\varrho  _{+}\equiv \varrho  (\bbox{x})=\rho (\bbox{x})-\rho _{0}$; $\varrho  _{-}(\bbox{x})=\rho ^{1}(\bbox{x})-\rho ^{2}(\bbox{x})-\rho ^{1}_{0}+\rho ^{2}_{0}$ with $\rho ^{1}(\bbox{x})=\rho ^{1\upA }(\bbox{x})+\rho ^{1\dnA }(\bbox{x})$ 
and $\rho ^{2}(\bbox{x})=\rho ^{2\upA }(\bbox{x})+\rho ^{2\dnA }(\bbox{x})$.  The Coulomb energy $E_{C}^{+}$, possessing the SU(4) 
symmetry, is the driving force to realize the QH system.  The term $E_{C}^{-}$ 
describes the capacitive charging energy between the two layers.  It vanishes 
in the limit $d\rightarrow 0$.  

The Hilbert space $\HilbLLL$ is defined by the LLL condition 
(\ref{LLLcondiDress}).  In the semiclassical approximation the wave function is 
given by (\ref{FactoWave}) at $\nu \leq 1$.  The SU(4) spin-pseudospin coherence is shown 
to develop spontaneously, precisely as the SU(2) spin coherence is.  The 
ground state at $\nu \leq 1$ is given by \cite{EzaIQC}
\begin{eqnarray}
\langle \varphi ^{1\upA }\rangle _{\txg}&&= \sqrt {\rho _{0}/2}\sqrt {1+\sigma _{0}}, \quad  \langle \varphi ^{1\dnA }\rangle _{\txg}=0, \nonumber\\
\langle \varphi ^{2\upA }\rangle _{\txg}&&= \sqrt {\rho _{0}/2}\sqrt {1-\sigma _{0}}, \quad  \langle \varphi ^{2\dnA }\rangle _{\txg}=0, 
\label{GrounOneLayer}
\end{eqnarray}
where $\rho ^{1}_{0}={1\over 2}\rho _{0}(1+\sigma _{0})$ and $\rho ^{2}_{0}={1\over 2}\rho _{0}(1-\sigma _{0})$.  It is convenient to use a 
new set of CB fields,
\begin{eqnarray}
\varphi ^{S\upA }&&= \sqrt {(1+\sigma _{0})/2}\varphi ^{1\upA } + \sqrt {(1-\sigma _{0})/2}\varphi ^{2\upA }, \nonumber\\
\varphi ^{A\upA }&&= \sqrt {(1-\sigma _{0})/2}\varphi ^{1\upA } - \sqrt {(1+\sigma _{0})/2}\varphi ^{2\upA }, 
\end{eqnarray}
and a similar set for the spin-down fields.  They are reduced to the symmetric 
and antisymmetric fields at the balanced point ($\sigma _{0}=0$).  We call them the 
``bond'' and ``antibond'' fields.  The ground-state value (\ref{GrounOneLayer}) is 
transformed into $\langle \widetilde{\Phi }\rangle _{\txg}=\sqrt {\rho _{0}}(1,0,0,0)$ in terms of the new 
fields, $\widetilde{\Phi }\equiv (\varphi ^{S\upA }, \varphi ^{S\dnA }, \varphi ^{A\upA }, \varphi ^{A\dnA })$.  

The tunneling energy is
\begin{equation}
E_{T} = -{1\over 2}\DSAS \sqrt {1-\sigma _{0}^{2}}\int d^{2}x [\varrho  _{-}^{S}(\bbox{x})-\varrho  _{-}^{A}(\bbox{x})] ,
\label{TunneEnergA}
\end{equation}
where $\varrho  _{-}^{S}(\bbox{x})=\rho ^{S}(\bbox{x})-\langle \rho ^{S}(\bbox{x})\rangle _{\txg}$ with $\rho ^{S}=\varphi ^{S\upA \ddag }\varphi ^{S\upA }+\varphi ^{S\dnA \ddag }\varphi ^{S\dnA }$ and similar 
equations for $\varrho  _{-}^{A}(\bbox{x})$.  The tunneling gap is $(1-\sigma _{0}^{2})^{1/2}\DSAS$ on the state 
(\ref{GrounOneLayer}).

\begin{figure}[thb]
\epsfxsize=70mm
\epsfbox{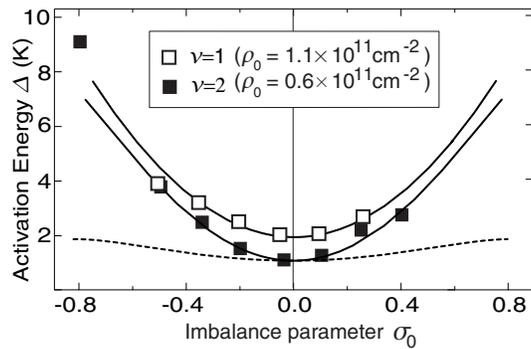}\caption{
Theoretical curves versus experimental data due to Sawada et al.[6].  The 
skyrmion charge is $e$ at $\nu =1$ and $2e$ at $\nu =2$.  The discrepancy of the 
data for large $|\sigma _{0}|$ may indicate that genuine CP$^{3}$ skyrmions are excited 
there since the tunneling gap $(1-\sigma _{0}^{2})^{1/2}\DSAS$ becomes smaller.  The dotted 
curve is for a would-be skyrmion carrying $e$ at $\nu =2$.  }
\label{ExcEneB1PS}
\end{figure}

A skyrmion excitation flips in general spins and pseudospins.  It is a 
CP$^{3}$ skyrmion \cite{Dadda} described by
\begin{equation}
\langle \widetilde{\Phi }\rangle _{\sky}=\sqrt {\rho _{0}}(z,\kappa _{1},\kappa _{2},\kappa _{3}) ,
\label{SkyrmOneBL}
\end{equation}
with constant parameters $\kappa _{i}$.  Its classical configuration is determined by 
(\ref{PreSolitEq}) $\sim $ (\ref{SkyrmCharg}), and (\ref{SkyrmDensi}) with $\kappa ^{2}=\sum _{\alpha }\kappa _{\alpha }^{2}$.  For 
definiteness we assume hereafter that the tunneling energy is much larger than 
the Zeeman energy.  (For instance, $\DSAS/g^{*}\mu _{B}B\simeq 4$ at $B=$ 5 Tesla in the 
sample of Ref.\cite{SawaPRLa}.)  Then, we have $\kappa =\kappa _{1}\not=0$, $\kappa _{2}=\kappa _{3}=0$.  It is 
identical to the CP$^{1}$ skyrmion (\ref{SkyrmSpin}) in the spin space.  At the 
balanced point the skyrmion size, the skyrmion energy and the flipped-spin 
number are given by the same formulas as (\ref{OptimScale}) $\sim $ (\ref{FlippSpin}), where 
$\beta $ now depends on the layer separation $d$.  Here, we concentrate our 
attention to its dependence on the imbalance parameter $\sigma _{0}$.  The term 
involving $\sigma _{0}$ is only the charging term (\ref{HamilCouloBL}) for $\nu \leq 1$.  We give a 
numerical estimation of the activation energy at $\nu =1$ in Fig.\ref{ExcEneB1PS} 
by using the sample parameters ($d=231$\AA, $\ell _{B}=120$\AA) of Ref.\cite{SawaPRLa}.  
The theoretical curve explains the experimental data quite well with a 
reasonable skyrmion size $\kappa \simeq 1$.

\textit{BLQH ferromagnet at $\nu =2$}: 
A caution is needed to analyze the BLQH system at $\nu =2$ since one Landau state 
contains two electrons.  We attach one unit of flux to each electron and 
transform it into the CB field by formula (\ref{DressField}) with $m=1$.  The 
effective magnetic field is not given by (\ref{EffecMagne}) but by
\begin{equation}
2\B_{\eff}=2B_{\perp } - \varepsilon _{ij}\partial _{i}\partial _{j}\Theta (\bbox{x}) = 2B_{\perp }-(2\pi /e)m\rho (\bbox{x}) .
\end{equation}
It vanishes on the homogeneous ground state at $\nu =2/m$.  Due to the fermi 
statistics the wave function is the antisymmetric product of two wave 
functions (\ref{FactoWave}),
\begin{eqnarray}
\F[\bbox{x}]=&& \prod _{r}[\langle \Phi _{1}(\bbox{x}_{r})\rangle \otimes \langle \Phi _{2}(\bbox{x}_{r})\rangle -\langle \Phi _{2}(\bbox{x}_{r})\rangle \otimes \langle \Phi _{1}(\bbox{x}_{r})\rangle ] \nonumber\\
&& \times \Laugh^{2}, 
\label{FactoWaveTwo}
\end{eqnarray}
where $\langle \Phi _{1}(\bbox{x})\rangle $ and $\langle \Phi _{2}(\bbox{x})\rangle $ are analytic and satisfy
\begin{equation}
\sum _{\alpha }|\langle \varphi ^{\alpha }_{1}(\bbox{x})\rangle |^{2}=\sum _{\alpha }|\langle \varphi ^{\alpha }_{2}(\bbox{x})\rangle |^{2} ,
\label{ExtraCondiTwo}
\end{equation}
as follows from the semiclassical LLL condition (\ref{PreSolitEq}).

When $\DSAS\gg g^{*}\mu _{B}B$, the spin-up and spin-down ``bond'' states are 
filled.  Hence, the ground state is given by (\ref{FactoWaveTwo}) with a set of two 
constant CB fields,
\begin{equation}
\langle \widetilde{\Phi }_{1}\rangle _{\txg}=\sqrt {\rho _{0}\over 2}(1,0,0,0),\quad 
\langle \widetilde{\Phi }_{2}\rangle _{\txg}=\sqrt {\rho _{0}\over 2}(0,1,0,0),
\label{GrounStateTwo}
\end{equation}
in terms of the ``bond'' and ``antibond'' fields.  This might be identified 
with the canted state \cite{DasSarma} for $\DSAS\gg g^{*}\mu _{B}B$.  A skyrmion excitation 
flips pseudospins, or induces a coherent tunneling excitation.  It is 
described by (\ref{FactoWaveTwo}) with a set of two CB fields,
\begin{eqnarray}
\langle \widetilde{\Phi }_{1}\rangle _{\sky}&&= \sqrt {\rho _{0}/2}(z,\kappa _{1},\kappa _{2},\kappa _{3}),\nonumber\\
\langle \widetilde{\Phi }_{2}\rangle _{\sky}&&= \sqrt {\rho _{0}/2}(\kappa '_{1},z,\kappa '_{2},\kappa '_{3}) ,
\label{SkyrmTwoBL} 
\end{eqnarray}
with $\kappa ^{2}\equiv \sum _{\alpha }\kappa _{\alpha }^{2}=\sum _{\alpha }{\kappa '}^{2}_{\alpha }$.  It consists of two CP$^{3}$ skyrmions (\ref{SkyrmOneBL}), 
and the skyrmion charge is $2e$.  We emphasize that there exists no skyrmion 
composed of one CP$^{3}$ skyrmion at $\nu =2$ because of the constraint 
(\ref{ExtraCondiTwo}).

An estimation of the excitation energy of the skyrmion (\ref{SkyrmTwoBL}) is 
straightforward.  We concentrate our attention to its dependence on the 
imbalance parameter $\sigma _{0}$.  The terms involving $\sigma _{0}$ are the charging energy 
(\ref{HamilCouloBL}) and the tunneling energy (\ref{TunneEnergA}).  The charging energy 
increases while the tunneling energy decreases as $\sigma _{0}$ increases.  We give a 
numerical estimation in Fig.\ref{ExcEneB1PS} by using the sample parameters 
($d=231$\AA, $\ell _{B}=228$\AA, $\DSAS=$6.8K) of Ref.\cite{SawaPRLa}.  The vortex limit 
($\kappa \simeq 0$) gives a best fitting of the experimental data because of a large 
tunneling gap ($\DSAS=$6.8K).  We have also given a theoretical curve for a 
would-be skyrmion carrying charge $e$ at $\nu =2$ by using the same parameters, 
where the charging energy and the tunneling energy are found to cancel each 
other almost completely (Fig.\ref{ExcEneB1PS}).  

The driving force of the SU(4) spin-pseudospin coherence is the Coulomb 
exchange energy arising from the SU(4)-invariant Coulomb term $E_{C}^{+}$ in 
(\ref{HamilCouloBL}).  Provided the exchange energy is dominant, it is obvious that 
the SU(4) coherence develops also at $\nu =2/m$ with the skyrmion charge $2e/m$, 
and at $\nu =6,10,14,\cdots $ with charge $2e$.

I am very grateful to Kenichi Sasaki, Anju Sawada, Sankar Das Sarma, 
Eugene Demler, Allan MacDonald and Jim Eisenstein for fruitful conversations 
on the subject.  Part of this work was done at ITP, Santa Barbara.  This 
research was supported in part by the National Science Foundation under Grant 
No. PHY9407194.


\end{document}